\documentclass{article}
\usepackage[utf8]{inputenc}
\usepackage[english]{babel}
\usepackage{amssymb,amsfonts,bm,latexsym}
\usepackage{float}
\usepackage[tbtags]{amsmath}
\usepackage{comment}
\usepackage{graphicx}
\graphicspath{ {./images/} }
\usepackage{authblk}
\usepackage[authoryear]{natbib}
\usepackage{csquotes}

\title{How A/B testing changes the dynamics of information spreading on a social network}
\author[1,2]{Matteo Ottaviani\footnote{Corresponding author, email: \texttt{matteo.ottaviani@sns.it}}}
\author[2]{Pietro Leonardo Nickl}
\author[2]{Stefan M. Herzog}
\author[2]{Philipp Lorenz-Spreen}

\date{Mar 12, 2021 }

\affil[1]{\small{Scuola Normale Superiore, Piazza dei Cavalieri 7, 56126 Pisa - Italy}}
\affil[2]{\small{Max Planck Institute for Human Development, Lentzeallee 94, 14195 Berlin - Germany}}

\begin{document}

\maketitle

%\tableofcontents

\begin{abstract}
A/B testing methodology is generally performed by private companies to increase user engagement and satisfaction about online features. Their usage is far from being transparent and may undermine user autonomy (e.g. polarizing individual opinions, mis- and dis- information spreading). For our analysis we leverage a crucial case study dataset (i.e. Upworthy) where news headlines were allocated to users and reshuffled for optimizing clicks. Our center of focus is to determine how and under which conditions A/B testing affects the distribution of content on the collective level, specifically on different social network structures. In order to achieve that, we set up an agent-based model reproducing social interaction and an individual decision-making model. Our preliminary results indicate that A/B testing has a substantial influence on the qualitative dynamics of information dissemination on a social network. Moreover, our modeling framework promisingly embeds conjecturing policy (e.g. nudging, boosting) interventions.
\end{abstract}

\section{Introduction}

%A review  of  the  relevant  literature that  motivates  the  research  question

The online ecosystem has been an exceptional ground for private companies to build choice environments that draw people's attention in order to make and increase profits. Although the their goal may vary from influencing political opinions to selling a pair of shoes, the methods which are involved often share the same algorithmic principles. One example in that respect is A/B testing, a method that optimizes pieces of the online environment to some goal function via a randomized experiment with two variants, the currently used version A (control) and its modification in some respect B (treatment).
%By testing a subject's response to A against B, this methodology aims to determine which one is more effective in optimizing to the set goal function.
%Typical ways to validate result may be either through the "frequentist" two-sample hypothesis testing or the "bayesian" one of probability distribution comparison, assuming the experiment outcomes derived from a Bernoulli trial. 
A/B tests are generally performed by private companies to increase user engagement and satisfaction about online features. Large platforms like Facebook, Instagram, Google, Groupon, LinkedIn, Microsoft, Netflix, Yahoo and Amazon use A/B testing to make user experiences more profitable and as a way to streamline interface of their services \citep{kohavi2014seven,xu2015infrastructure}. Surfing the web, users undergo A/B tests without awareness; for example different users may see different versions of the same web page at the same moment.   
Thereby, the A/B testing machinery is far from being transparent; private companies don't share their data, their methods, their direct purposes and, particularly, the different versions they are testing.

We may discern two different levels of impact: the way A/B testing influence individual decision making and the way how it scales up to collective behavior. 
The first research question that we will address here aims to find out which are the most relevant features that A/B testing is typically amplifying. In order to do that, we will leverage real world data from a particular field of application of A/B testing. In the second research question we will ask how and under which conditions A/B testing affects the distribution of content on the collective level, specifically on different social network structures.
Understanding the impact A/B testing has on the online ecosystem and the mechanisms at work may help to develop intervention and tools that dampen potentially unwanted outcomes of A/B testing and help to promote ``user autonomy'' \citep{lorenz2020behavioural,kozyreva2019citizens}.

%a full description  of  the  experimental  aims  and  hypotheses

We will leverage a crucial case study regarding the effectiveness of A/B testing: the Upworthy archive \cite{matias2019upworthy}. In November 2019 Good \& Upworthy and a team of researchers announced a dataset of 32,487 A/B tests carried out by Upworthy from January 2013 to April 2015 \citep{matias2019upworthy}. The company conducted its A/B tests on its own website, randomly varying different presentations for a single story. 
Presentations were \emph{packages} of news/story headlines that were randomly assigned to people on the website as part of a test. A/B test machinery compared package ``success'' to pick winners employing the click-through rate (CTR; i.e. the ratio of clicks to impressions).
The present data set is by definition a clickbait dataset: The different headlines where designed and tested in an effort to maximise clicks \citep{BLOM201587,chen2015misleading}. 
This data is particularly suitable for systematically analyzing which linguistic features successfully attract clicks and are then amplified through A/B testing.

Linguistic cues occur on different levels. Particular attention in each package should be paid on the lexical or semantic level, the structural or grammatical one and the formal or formatting one. For example, different words may recall different emotional reactions, some topics could have been more successful than others, a specific text and word lengths may impact in a certain way.

We will combine rule-based and machine learning methods to extract linguistic features. For the semantic analysis we will use a topic model and a text analysis tool that incorporates a number of freely available sentiment dictionaries. 
We will use topic modeling to detect the presence of semantic topics (e.g. feminism, racial equality, LGBT+ issues, etc). For the topic modelling we will use Latent Dirichlet Allocation (LDA). LDA is an unsupervised learning algorithm that detects the co-occurence of words across documents (e.g. a headline) in a corpus (the collection of all headlines). Words that cluster together form a ``topic'', which may or may not be an interpretable semantic topic. 
In order to detect the presence of emotion, arousal etc. we will use Sentiment Analysis and Cognition Engine \citep[SEANCE;][]{crossley2017sentiment}, a freely available text analysis tool that embeds a number of sentiment dictionaries.
Mainly we are using SEANCE to analyse headlines in terms of valence, arousal and specific emotions. 
Once we obtained linguistic features characterizing packages, we associate to each package an input vector made of its extracted features. 
The decision-making model $DM$ we assume is a simple linear relationship that binds the linguistic features present in a generic package with its probability to be clicked, i.e., the click-through rate of that package. 
We assume that the decision-making model holds globally for each package of the Upworthy data set. The continuous outcome for each package (i.e. the click-through rate), together with the latter assumption, allows for a linear regression method. Therefore, we perform a LASSO regression (Least Absolute Shrinkage and Selection Operator). The regression coefficients coming from its training quantify the overall impact of each linguistic feature on click-through rate; they serve the decision-making function for simulating an individual interacting in front of an hypothetical (e.g. clickbait) headline.
Therefore, we put forward a data-driven human decision-making model for clicking headlines online, based on linguistic/semantic features as cues. 
%An agent is modeled as a click rate made up by linearly compounding the impact of linguistic features present in the package.

To asses the global impact of A/B testing on the distribution of content, we compare two situations: (1) clicks on headline refer to a social sharing process, thereby no global actor has an agency and just the attractiveness of linguistic features plays a role in the distribution and (2) a global A/B tester is introduced who is aiming to additionally maximize the clicks by systematically varying the headlines following an A/B testing scheme. 
%Let us consider the case of simulating an ideal social environment in which individuals interact sharing contents to their friends and acquaintances.  
The two main assumptions of our agent-based model lay on:
\begin{itemize}
    \item all agents are equal: each of them reacts to a given package following the same decision-making model (mentioned above);
    \item the click for an agent means ‘sharing’ a package (partially of entirely) to the nearest neighbors (with a friction parameter: the infection rate).
\end{itemize}

In order to mimic an online discourse we couple agents via a stylized social network structure, along which they can share information (headlines in this case).
We employ several network topologies (e.g. Albert-Barabasi, Erdős-Rényi, Stochastic Block Model) and tune their densities ( i.e. sparsity of the network).

We consider synthetic randomly generated headline, which consists of a random combination of a fixed number of linguistic features.

%In order to pursue our second research question, i.e. under which conditions a piece of news enhanced by A/B testing spread in a social network structure, we need to set up a general framework in which we can ideally 'switch on' the A/B testing machinery and perform a comparison afterwards. Then, we design the agent-based model and we can distinguish between two scenarios. In the first scenario, messages shared by individuals undergo a pure social spreading; in the second one, instead, A/B tests lead the message selection dynamics.
In a pilot studies we have tested the procedure mentioned above. It has been achievable thanks to the availability of an exploratory data set from the Upworthy Research Archive \cite{matias2019upworthy}.
The preliminary results coming up from the exploration of the exploratory data set allowed us to generate research questions that we wish to address with the full data. 
In particular, the final distributions of feature distribution from our agent-based model, either in a pure social spreading scenario or in the A/B testing condition. We could observe that the A/B testing mechanism increases the homogeneity of information that is spread. In other words, we observe that A/B testing performed on a social network structure reduces the exploration and amplifies exploitation of successful features of early pieces of information, ignoring others. With the full data from the Upworthy archive, we will be able to answer what types of information this typically is and is not.

%------
\section{A/B testing: a brief overview}

Private companies which sell their products online, including Amazon, eBay, Etsy, Facebook, Google, Groupon,  LinkedIn, Microsoft, Netflix or Yahoo \citep{kohavi2014seven,xu2015infrastructure}, have been using \textit{online controlled experiments} (i.e. A/B testing) for at least two decades, in order to shape their platforms. However, in their words, the aim that drives them is the continuous improvement of users' online experiences.
Companies run and analyse thousands concurrent experiments per day in order to validate new ideas, which materially translate in changes in the online environment shown to customers. The latter range from entire redesigns and infrastructure changes (e.g. a post on a social network, mobile app interface, etc) to bug fixes. Finally, the whole collection of successes and failures is then employed in learnings on customer behaviour \citep{kaufman2017democratizing,xu2015infrastructure}. 
Private companies are used to build in-house experiment infrastructures to maximize their products' success, strongly relying on A/B testing \citep{kaufman2017democratizing,xu2015infrastructure}.

As briefly mentioned before, A/B testing is a randomised controlled trial for assessing the causal effect of introducing a new idea (i.e. treatment) on some wished output. Traditionally, the effect of the treatment is estimated on the whole intended population, since the outcome for a given individual cannot be computed for both versions A and B. It is referred to as the Average Treatment Effect \citep[ATE;][]{rubin1974estimating}: it measures the difference in the randomized treatment and control groups as following: $ATE= E[Y=1]-E[Y=0]$. The ATE is then the difference in the expected values of the treatment and control group’s output. For a given causal effect of interest, the ATE value would suggest the effectiveness of some treatment applied. Statistical inference is usually employed to determine whether an ATE estimate is statistically consistent \citep[either positively or negatively;][]{kaufman2017democratizing}.
However, since ATE is a measure of the average effect, a positive or negative ATE does not assess whether and how a particular individual would react to a given treatment. In the last years, much more attention has been focused on the latter requirement. In particular, the rising of huge data sets containing personalized treatments, along with their individual profiles, has allowed to explore how treatment effects vary across individuals. 
%Researchers, businesses and policymaker have wished to overcome ATE in randomized experiments rather heading to how personalize treatments and to better understand causal relationships. 
Accordingly, a new measure has been put forward: Conditional Average Treatment Effect (CATE, $\tau(x)= E[Y(1)-Y(0)|X=x]$) which is the treatment effect conditional on observed covariates. Meta-algorithms are employed in order to compute CATE. The most common of them takes two steps: it employs the so called ``base learners'' to measure the conditional expectations of the outcomes both for control and treatment groups, separately; afterwards, it takes the difference between these estimates. To date, most used meta-algorithms include:\textit{T-learner}, \textit{S-learner} and \textit{X-learner} \citep{kunzel2019metalearners}.   

%------------------

\section{A case of study: Upworthy.com}

We took into account a crucial case study regarding the effectiveness of A/B testing: Upworthy Archive. Upworthy.com is a company founded in 2012 by the author of \textit{The Filter Bubble} \citep{pariser2011filter} and cofounder Peter Koechley; their aim was to reach large audiences with pieces of news (in their words ``stuff that matters'') which would have drawn most of people's attention on the web. 
A key component of their strategy was to setting up attractive ``packages,'' each of them consisting of a headline, a subheading and an image, similar to the one individuals were shown when an article was posted on a social media.
In order to establish which package would influence the most, Upworthy adopted A/B testing practices. They were already popular among technology industry and political campaigns. The company optimized content to deliver packages, measure responses, and compute probability of a viewer clicking on different versions of the same story. In the history of American media, Upworthy was a leading actor from 2013--2015; people associated its success to the idea of ``clickbait.''
Editors and website engineers reported that they set only one experiment per page in order to reduce correlations and/or dependencies among experiments.
The system recorded the number of individuals that were shown a given package (i.e. \textit{impressions}) and the number that clicked on the package (i.e. \textit{clicks}). After a while, editors decided either to choose the best performing package for finalizing or to keep modifying packages for increasing performance \citep{matias2019upworthy}. In November 2019 Good \& Upworthy (Upworthy merged with Good Worldwide in 2017) and a team of researchers composed by J. Nathan Matias (Cornell University), Kevin Munger (Penn State University), and Marianne Aubin Le Quere (Cornell University) announced \textit{The Upworthy Research Archive}, a dataset of 32,487 A/B tests carried out by Upworthy from January 2013 to April 2015. In their words, the release of the dataset aimed to enhance knowledge in many fields, including: communication, political science, psychology, statistics, and computer science \citep{matias2019upworthy}.

\subsection{The Upworthy Research Archive}

\begin{figure}[h!]
\centering
\includegraphics[scale=0.28]{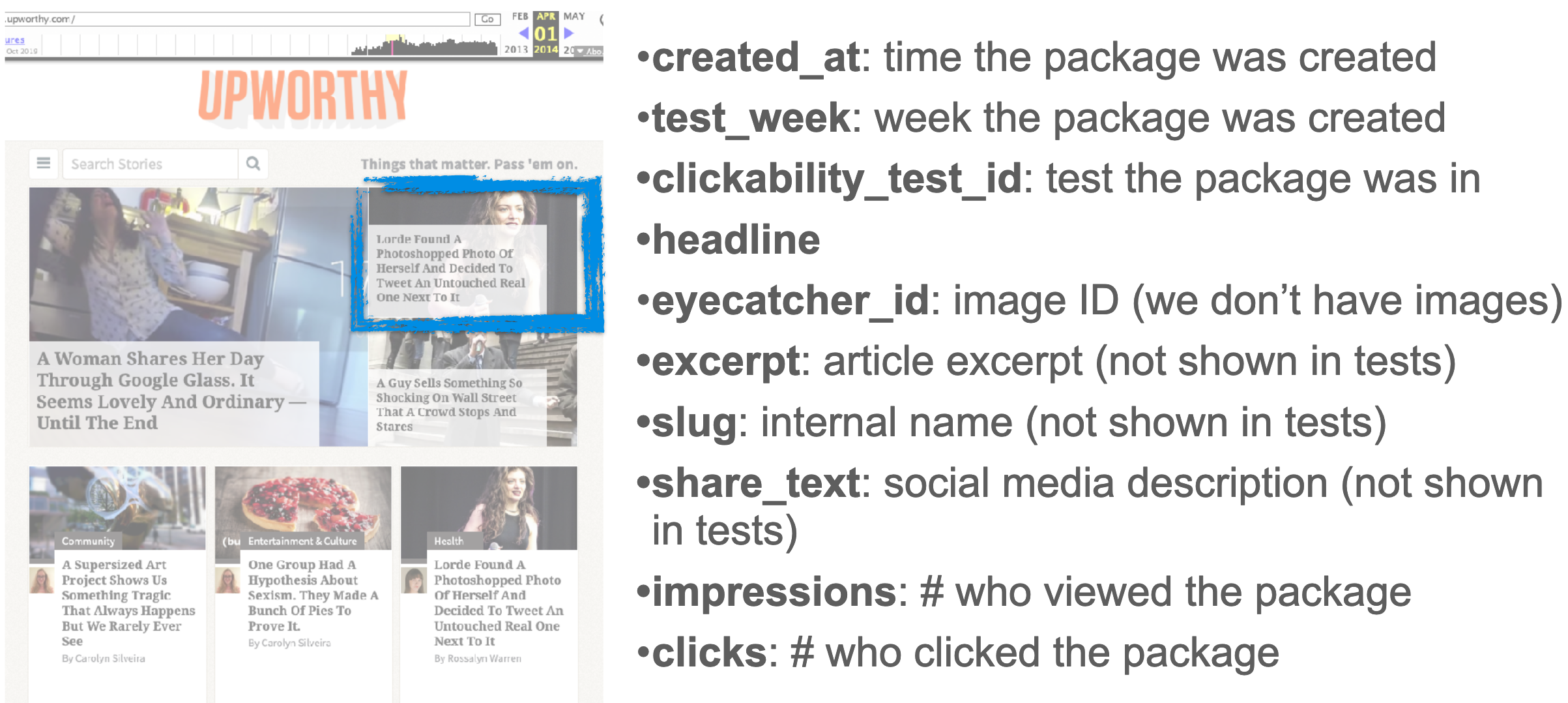}
\caption{An example of package features from the Upworthy Research Archive, as described in \cite{matias2020asking}.}
\label{fig:upworthy_example}
\end{figure}

The Upworthy Research Archive dataset consists of 32,488 tests carried out from January 24, 2013 through April 14,  2015,  just after an editorial shifts was announced by the company. For each test, the dataset includes viewer reactions to each package in an experiment; there is a median of 4 packages per test. The dataset contains over 150 thousand packages. These packages altogether received over 538 million impressions and over 8 million clicks. Each test includes a median of 14,342 impressions and a median of 201 clicks per test.
Each package consists of: the experiment ID; the headline; the subhead, the social media summary (where Upworthy used them); the preview image; the number of impressions and clicks received by the package during the test \citep[see Fig.~\ref{fig:upworthy_example};][]{matias2019upworthy}.

\vspace{5mm}

\subsection{Clickbait: definition and characteristics, linguistic features}

\cite{potthast2018clickbait} distinguish between four definitions of clickbait by: emergence, intention, effect and perception. We may define clickbait by its effect (as whatever attracts the most clicks) or as a certain style (e.g. anything being perceived as sensationalist). A New York Times headline might be very effective without employing clickbait style. It is worth distinguishing between clickbait in the effective definition and ``clickbait style.'' It could be that clickbait style actually is not the most effective way of grabbing a reader’s attention, or that it even annoys some people---for example the makers and users of downworthy (ZOT), a browser extension that changes clickbait style elements, replacing words like ``Awesome'' and ``WHOA.'' While we expect our present dataset to fulfil all four definitions of clickbait, it is important to keep in mind the distinction between clickbait as a certain style and clickbait as a click-optimised piece of text. 

\subsection{Clickbait style}

The present data set is by definition a clickbait dataset: It is the result of an effort to optimize for number of clicks. Though the effectiveness of any headline may not be due to clickbait style, it is worth examining this style in the context of the Upworthy dataset, as it actually coined much of what we think of as ``clickbait style.'' In fact, studies on clickbait actually take material from upworthy \citep{BLOM201587,chen2015misleading}. Clickbait style includes formal features (the two-sentence headline, the listicle, unusual punctuation) as well as semantic features (colloquial, overly emotional language). We cast a wide net: We try to include as many clickbait characteristics from the literature as possible. 

\section{Methods}
% genral description and graphical rendering

\begin{figure}[h!]
\centering
\includegraphics[scale=0.18]{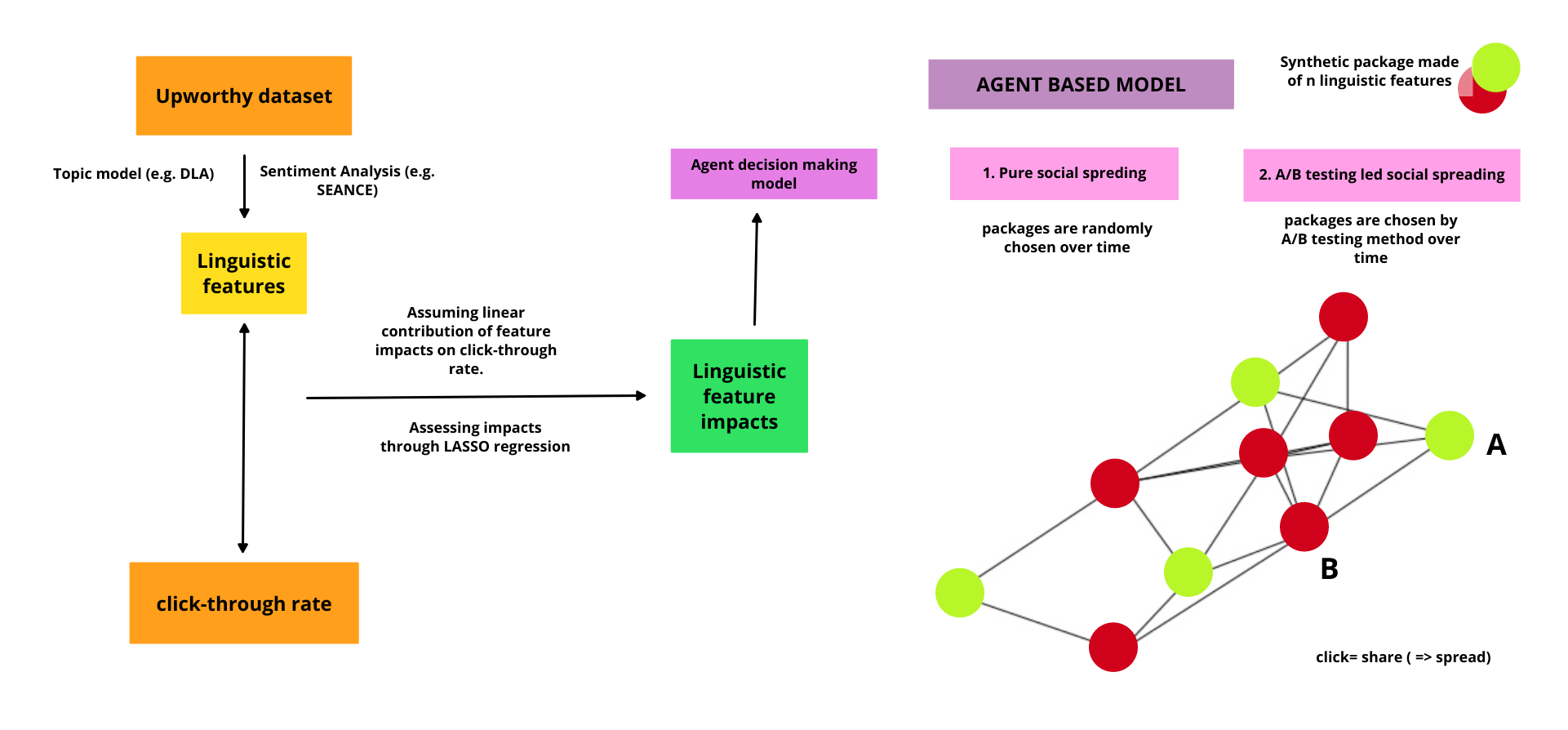}
\caption{Broad round-up of the temporal order of our procedure.}
\label{fig:schematicprocedure}
\end{figure}

In Fig.~\ref{fig:schematicprocedure} we show a broad outline of the main steps of the procedure we follow.
After employing linguistic methods on the data set in order to infer linguistic features, we convert our data set in input-output relationships for each package, where the features presented in each bundle are linked to its click-though rate. By performing LASSO regression over the latter, we infer feature coefficients which serve us their overall impacts (negatively or positively) on the click-though rate.
Our first research question about what are the linguistic features most mattering in an A/B testing can be addressed with this procedure. 
For the second research question we will assume a simple linear model for decision making that binds the linguistic features present in a generic package with its probability to be clicked, directly linking to the click-through rate of that package. LASSO regression coefficients are part of this individual decision-making model to quantify the feature impacts.
In order to pursue our second research question, i.e. under which conditions a message enhanced by AB testing spread in a social network structure, we need to set up a general framework in which we can ideally 'switch on' the A/B testing machinery and make a comparison afterwards. Therefore, we design a simplistic agent-based model, where we can distinguish between two scenarios. In the first one, messages shared by individuals undergo a pure social spreading; in the second one, instead, A/B tests influences the message selection dynamics.
In the following we go through a full description of all methods that we will employ for achieving this.

\vspace{5mm}

\subsection{Linguistic features extraction}

We combine rule-based and machine learning methods to extract formal and semantic linguistic features. Most linguistic features are extracted by the application of an explicit rule, e.g. the presence of a certain character or word. For the topic modelling, we employ a data-driven approach, by capturing the underlying semantic structure of the corpus. All code is implemented in python. 
For preprocessing we employ a separate pipelines based on the python libraries nltk, and spaCy separately, as they come with different functions for tokenization, lemmatization and different word lists e.g. for stop word recognition.
The output of the feature extraction process is a pandas dataframe. It is a headline-feature matrix with the headlines as rows and the extracted linguistic features, as well as the click-rate as columns. We will use this matrix both to describe the upworthy corpus linguistically as well as to see how features are associated with the click rate.

\subsubsection{Formal Features}
Clickbait style is characterised by formal features, such as a preference for fully formed sentences as titles. This entails overall more words, a higher stop-word to content-word ratio and longer syntactic dependencies in comparison to the succinct headlines of established newspapers, while average word length might be low. Such features are easy to operationalise. For our feature extraction, we take inspiration from \cite{kuiken2017effective} and \cite{zheng2017boost} who list these and more features (see table 1 for their respective lists). As formal features \cite{safran2013data} lists five headlines types: normal, Question (headline forms a question), How to (headline starts with ``how to''), Number (headline introduces a listicle), and reader-addressing (contains a form of ``you''). We would like to include the feature of forward referencing \citep{BLOM201587}, another characteristic of clickbait style: In headlines like ``She Did Not Expect THIS,'' or ``What Happened Next Will Blow Your Mind.'' Here, one has to click the article to resolve what these headlines are referring to. We are still working on an adequate operationalisation of this feature, perhaps simplifying it to the presence of a demonstrative and/or pronoun (like this, she). 

\begin{table}[h!]
\centering
\begin{tabular}{p{0.38\linewidth}p{0.38\linewidth}}
\hline
\cite{zheng2017boost} & \cite{kuiken2017effective}  \\
\hline
Presence/number of exclamation mark & Number of characters  \\
Presence/number of question mark & Number of words           \\
Presence of pronoun & Average word length \\
Presence of interrogative &  Number of sentences  \\
Number of words & Number of sentimental words  \\
Number of dots & Readability score  \\
Stop words ratio & Containing question \\
tf-idf weight of words & Containing quote \\
tf-idf weight of bigrams & Containing signal words \\
 & Containing pronouns \\
 & Containing number \\
 & First word type \\
\hline
\end{tabular}
\caption{Clickbait feature lists to compute a clickbait score.}
\end{table}

%SEMANTIC FEATURES – SENTIMENT ANALYSIS AND TOPIC MODELLING 
\subsubsection{Semantic features}
%SENTIMENT ANALYSIS
\paragraph{Sentimental Analysis} 

Emotional valence (positive/negative) and arousal (strong/weak) are important factors for reading behaviour (Chen et al., 2015). Clickbait is further associated with overly positive sentiment. For sentiment analysis we will use the Sentiment Analysis and Cognition Engine (SEANCE, Crossley et al. 2017). SEANCE is a large knowledge base including many word lists for nuanced sentiment analysis (251 core indeces). Since SEANCE includes word lists that are interpretable as a topic (e.g. ``politics, economics, and religion,'' ``Dominance, respect, money, and power,'' ``social relations,'' ``Arts and Academics''), it may also serve as a knowledge base to detect these topics.\\

\paragraph{Topic model} We will employ topic modeling to detect the presence of semantic topics (e.g. feminism, racial equality, LGBT+ issues). For the topic modelling we will use Latent Dirichlet Allocation (LDA). LDA is an unsupervised learning algorithm that detects the co-occurence of words across documents (e.g. a headline) in a corpus (the collection of all headlines). Words that cluster together form a ``topic,'' which may or may not be an interpretable semantic topic. 
For the LDA, we will define a document as each unique story plus all its different headlines, i.e. we combine the unique “lede” with all its unique headlines. This yields larger, more semantically coherent documents, which give LDA a better chance of picking up semantically related words. This means that we will perform LDA on the story-level, not the headline level. 
%We could have performed LDA on the headline-level, taking each individual headline as a separate document, but in our preliminary analysis this did not yield interpretable topics. 
Since we aim to detect semantic topics, preprocessing for the LDA will include lowercasing, lemmatization, stopword removal, punctuation removal. While we may lose potentially informative formal features by removing formal features as ``noise,'' we will amplify the semantic signal. We will retrieve information about stopwords, uppercasing and punctuation in other feature extraction steps. 

\subsection{Assessing linguistic features importance in the data set}

Through the linguistic methods mentioned above, we will obtain in total $M$ linguistic features extracted along with their relative weights. In other words, for each package $k$ and for each feature $i$ included in it, a relative weight $F_i^k$ is measured. Each package $k$ identifies an online environment and it can be formally described as follows: $\{F_1^k,...,F_M^k \}$, where $F_i^k$ is the weight assigned by linguistic methods to the linguistic feature $i$ for the package $k$; $F_i^k=0$ if the linguistic feature $i$ was not found in the package $k$.

%$OE_k=\{w_1^k \sigma_1^k T_1,...,w_M^k \sigma_M^k T_M \}$, where $\sigma_i^k=1$ if the linguistic feature $i$ was found in the package $k$, $\sigma_i^k=0$ otherwise. 

\subsubsection{Individual decision-making model}

Our main purpose is to build up a decision-making model able to surrogate the interaction of an individual with a news headline presented online. The latter may be thought of consisting of a set of linguistic features: $OE_k=\{j,n,...,s \}$, where $j,n,...,s$ are the indices belonging to the linguistic features present in the headline $k$; $k=1,...,K$, $K$ is the total number of packages available. Let us call $Y_k$ the click-through rate value for package $k$.

The decision-making model $DM$ we assume is a simple linear relationship that binds the linguistic features present in a generic package $k$ with  the click-through rate of that package. Formally:  
\begin{equation}
    DM_k = \sum_{i \in {OE_k} } w_i = Y_k \hspace{10mm} \propto p_k 
    \label{eq:DM}
\end{equation}
where $w_i$ are the coefficients quantifying the overall impact of feature $i$ and $p_k \in [0,1]$. 

In order to assess features importance $w_i$'s, let us assume that our decision-making model (eq.\ref{eq:DM}) holds for each package of the Upworthy data set, that is a simple linear relationship which binds the linguistic features present in the package with its click-through rate. Since we are provided of a continuous outcome for each package (i.e. the click-through rate), together with the latter assumption, then a linear regression method fits these requirements.

\subsubsection{LASSO}

In order to measure parameters for all features, we will perform a Lasso regression (Least Absolute Shrinkage and Selection Operator). It is a modification of linear regression. In Lasso, the loss function is modified to minimize the complexity of the model by limiting the sum of the absolute values of the model coefficients (also called the $l1-$norm). 

Formally, the estimates of LASSO coefficient $(w_1,...,w_M)$ are the quantities that minimize:
\begin{equation}
    \sum_{k=1}^K Y_k - \left( w_0 + \sum_{i=1}^{M} w_i F_i^k\right)^2 + \lambda \sum_{i=1}^{M} |w_i| 
\end{equation}
where $w_0$ is the regression intercept and $\lambda$ is a tuning parameter which to be determined apart. The LASSO regression searches for the best coefficient estimates that fit the data set.

LASSO differs form other regulation methods in this context (as Ridge, OLS, etc) thanks to its second term, which is called ``shrinkage penalty;'' it has the effect of shrinking the coefficient estimates towards zero.
While
several regression methods tend to generate a model involving all features, the LASSO will not only shrink parameter estimates but it will also push some of them to zero. We could say that LASSO performs variable selection, in a certain sense \citep{georges2019market}.

\subsubsection{LASSO tuning parameter selection}
LASSO implementation needs a way to selecting the tuning parameter $\lambda$. Cross-validation (CV) provides a simple method to achieve this task and it is consistent with the goal of minimizing overfitting. 
In particular, we will employ $k$-fold CV that consists in randomly splitting the set of observed data into $k$ folds of equal size. The first fold works as a validation set, and the LASSO regression is performed on the remaining $k-1$ folds. 
The fold kept out (i.e. the validation set) is used to compute the mean squared error, $MSE_1$. By iterating this procedure $k$ times with a different fold of data set used as the validation set, the process generates $k$ estimates of the test error, $MSE_1, MSE_2,..., MSE_k$. The average of the latter yields the k-fold CV error:
\begin{equation}
    CV_k = \frac{1}{k} \sum_{j=1}^{k} MSE_j \,\,\text{.}
\end{equation}
The choices of $\lambda$ and $k$ are affected by a bias-variance trade-off. It has been shown empirically that $k=5$ and $k=10$ values produce test error rate estimates with relatively small bias and variance and it is then common practise to compute $k$-fold CV employing them \citep{georges2019market}.
In light of this, we will perform LASSO with respect to a grid of $\lambda$ tuning parameters, we will compute the cross-validation error for each of them, and then we will select the one for which the k-fold CV error is the lowest value.

Once the tuning parameter is chosen, we will finally obtain the best set of LASSO regression coefficients; they represent the overall impact of features on the click-through rate.

\subsection{Agent based model of messages social spreading}

Once obtained from LASSO performance the impact of each feature on the click-though rate, the decision-making function shown in Eq. \ref{eq:DM} is fully working for simulating an individual interaction in front of an hypothetical (e.g. clickbait) online environment.
Let us figure individuals in a social network which are given of different stimuli (in the spirit of Upworthy); their social interactions may promote sharing of popular messages (i.e. linguistic features). A synthetic reproduction of that dynamics may be thought as following. In our simulation frameworks, $N$ agents are given. Our first assumption lies on the fact that they are all equal: they behave through the same decision-making function (Eq. \ref{eq:DM}) in front of a package, provided of the same feature coefficients. The agents are linked one another with respect of a network structure which reproduce social interconnections among them. In order to mimic a social network structure, the best modeling tool which fits our requirements is borrowed from graph theory; we are endowed with a stylized complex network whose nodes are agents presented in the environment and whose edges are the possible interactions among different individuals.
We employ several network topologies (e.g. Albert-Barabasi, Erdős-Rényi, Stochastic Block Model) and tune their densities (i.e. sparsity of the network).
Let us consider synthetic packages generated by randomly picking $n$ linguistic features. When an agent is given of a package, its individual decision-making generates an output with respect to the features presented in the package. This response is then converted in a probability to click and a random threshold determines if the agent clicks or not on the content. Once an agent clicks, it \textit{shares} the successful package to its directly connected nearest neighbors, according to an infection rate $\eta in (0,1]$. The latter is our second main assumption: the click for an agent means ``sharing'' a package to the nearest neighbors (with a friction parameter $\eta$).
In order to pursue our second research question, i.e. under which conditions a piece of news enhanced by A/B testing spread in a social network structure, we need to set up a general framework in which we can ideally 'switch on' the A/B testing machinery and make a comparison afterwards. Therefore, we design the agent-based model and we can distinguish between two scenarios. In the first scenario, messages shared by individuals undergo a pure social spreading; in the second one, instead, A/B tests lead the message selection dynamics.

\subsubsection{Pure social spreading scenario}

The case of a pure social spreading is meant to be our benchmark scenario. The simulation starts by drawing two randomly generated packages (i.e. packages are done of $n$ linguistic features randomly picked) and are randomly allocated to agents all over the network, package $A$ to half population and package $B$ to the other half. Agents click or not according to our decision-making model and the ones who clicked may spread (according to the infection rate $\eta$) to those nearest neighbors of theirs who were shown the other package. This spreading dynamics stops in a few steps, once the opportunities of sharing a new content to neighbors saturate (i.e. each agent may at most see both packages and share them around if it clicked). At this point, a new two packages drawing takes place: one between the old $A$ and $B$ is kept and the other one is partially reassembled according to a mutation rate $\mu$; i.e. $\mu$ is the percentage of package features which is randomly varied. Another test administration round starts again by randomly splitting the population in two and randomly assigning the two packages to agents and so on. 

\subsubsection{A/B testing led social spreading scenario}

The case of social spreading in which A/B testing is performed by a third party (e.g. a private company) is a slight but pretty crucial modification of the benchmark framework mentioned above. The A/B testing machinery takes place from the second round package selection on. By comparing click results between $A$ and $B$ of the previous round, it chooses for the new test administration the $B$ package if its performance overcomes A click-though rate and builds a new package by varying $B$ of a mutation rate. In the opposite case, $A$ is kept for the new round and a new package $B$ is built by varying $A$ package. Exploring A/B testing literature, several statistical methods have been employed over time for assessing if $B$ click-through rate greater than $A$ one is statistically consistent in an A/B test. Nevertheless, the main difference is between \textit{frequentist} and \textit{Bayesian} approaches.

\paragraph{A/B testing methodologies}

In general, one performing an A/B test is basically collecting click-through rates from the control $A$ and the variation (B) and then uses a statistical method to determine which of the two performed better.
In a frequentist world, one would use p-values and make a choice between the null hypothesis (i.e. there is no relevant difference between $A$ and $B$ , then we keep $A$) and the alternative hypothesis that the variation $B$ is better than the control $A$. As soon as the p-value reaches statistical significance or large amount of data, the experiment is considered done. The frequentist approach requires a wider difference in performances in order to prefer $B$.

On the contrary, following the Bayesian methodology, the click-though rate of each variant is modeled as a random variable with a probability distribution. 
Even in case the improvement of a new variant is small, Bayesian statistics is more willing to choose the variation $B$ against the control $A$. On one hand, this means that this methodology results more prone to false positive rate; on the other hand, it lets either to control the magnitude of wrong decisions or to perform a quick series of experiments so that one may accumulate small gains from incremental improvements once making a lot of variations is effortless and free of charge \citep{frasco2018the}.

Since our aim is to simulate how a private company would perform A/B testing, our choice of which methodology to reproduce solely depends on which choice they would make. They actually depend on their purposes and their contexts. For example, one may prefer the latter in situations in which the treatment requires relevant expenses (e.g. engineering maintenance, disruption to the user experience, the costs of implementing it, etc) which would only be offset by large benefits (i.e. wide difference between $A$ and $B$ click-through rates). 
On the contrary, for situations similar to ours a Bayesian approach is rather more preferred. Each small improvement is very welcome since no additional expenses one incurs in choosing the variation $B$ against $A$, even if it is a false positive. It doesn't worth putting effort against the false positive rate; it is much more relevant administrate more tests (i.e. explore wider all the possible combinations) and quickly.

\subsubsection{Bayesian A/B testing}
The Bayesian methodology for performing A/B testing lies in a nicety of the Bayes theorem in combination with the nature of processes underlying A/B tests. 
It is usually very hard or impossible in most of cases to obtain closed form solutions to the Bayes theorem; for this reason approximation methods (e.g. Markov Chain Monte Carlo) have been developed.
An A/B test can be described by Bernoulli trial since it is a random experiment with only two possible outcomes: ``click'' and ``not to click.'' The Beta distribution is the conjugate prior of such process; that means the posterior function of the Bayes Theorem lies in the same family of the prior one and then a final function can be built with an iterative process. Therefore, an exact solution of the Bayes formula (i.e. closed-form solution) exists.
As just mentioned, a convenient prior distribution form modeling a binomial parameter $q$ is the beta distribution.
Starting from a flat, uninformative prior, defined by $Beta(1,1)$, the distribution of $p$ after $C$ clicks and $F$ failures (i.e. impressions minus clicks) is given by $Beta(C+1,F+1)$. $C+1$ and $F+1$ are the two parameters of the beta distribution for the belief.
Therefore, in our case we have two beta functions and accordingly their two Bayesian beliefs; they are one for the experimental branch $A$ and the other one for $B$:$p_A \sim Beta(C_A+1,F_A+1)$ and $p_B \sim Beta(C_B+1,F_A+1)$.

In line with \cite{miller2015formulas} and without explicitly report the calculation, the probability that $B$ will perform better than $A$ in the long run is given by:
\begin{equation}
    Pr(p_B>p_A)= \sum_{i=0}^{C_B-1} \frac{Beta(C_A+i,F_B +F_A)}{(F_B+i)Beta(1+i,F_B)Beta(C_A,F_A)} \,\,\text{.}
    \label{eq:prbgreatera}
\end{equation}
Let us define the \textit{uplift} as how much $B$ variation increases the click-through rate with respect to the control $A$:
\begin{equation}
    uplift= \frac{c_B - c_A}{c_A} \,\, \text{,}
\end{equation}
where $c_A$ and $c_B$ are click-through rates.
Let us suppose the latter is positive. In order to assess how trustful is the result, Eq. \ref{eq:prbgreatera} estimates the probability that $B$ option to perform better than $A$.
Note that in the frequentist approach there is no way to calculate this probability. In that case, one computes a p-value and in case it falls under an arbitrary threshold, one would state that ``within a certain percentage of confidence level (i.e. the $\alpha$ value, usually $95\%$), the null hypothesis can be rejected.'' This is a way different from the Bayesian case here where one would state ``this hypothesis is better then the other with a certain probability percentage.''

\vspace{5mm}
\begin{figure}[h]
\centering
\includegraphics[scale=0.3]{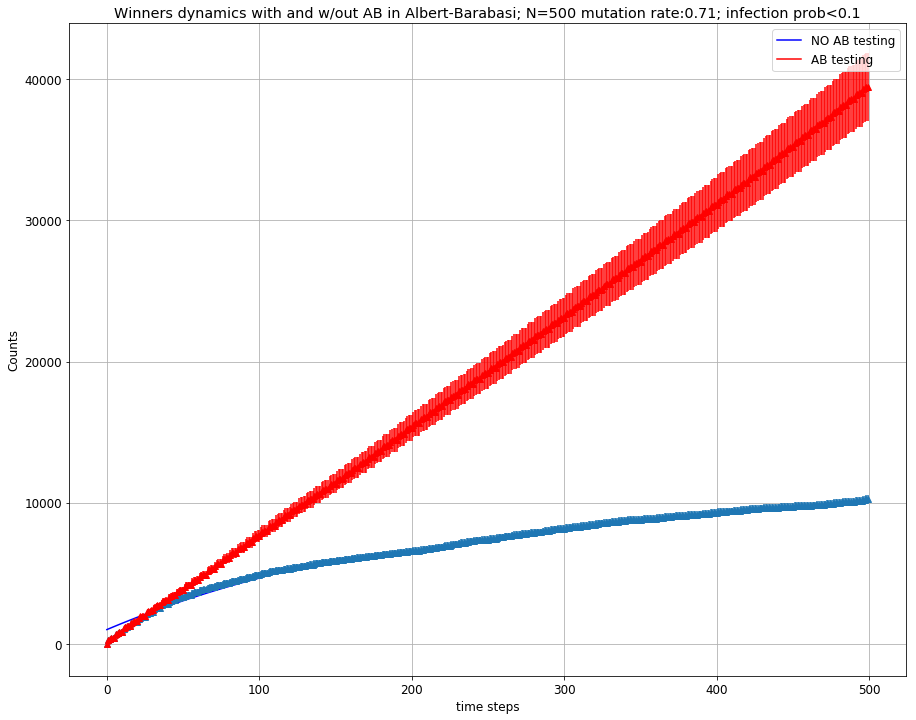}
\caption{Comparison of the fist-ranked linguistic feature dynamics in both explored scenarios, a pure social spreading setting and an A/B testing led one. The network structure employed in the simulation is the Albert-Barabasi one. }
\label{fig:successful}
\end{figure}

\section{Example of pilot data exploration}

All the procedure explained in details above has been tested exploratory data set from the Upworthy Research Archive \citep{matias2019upworthy} and we are happy to share the code either to replicate results or to perform new experiments. All the machinery described above may be considered pretty general and, in principle, it could employ successfully other similar data sets.    

In the following we present some results obtained from our procedure's performance. 
In particular, we set up a social spreading framework according to the settings previously described. We simulate $N=500$ agents allocated according to the Albert-Barabasi network structure for total $T=500$ time steps. Each $5$ time steps a new $A$-$B$ couple drawing takes place. We perform this simulation either the pure social spreading framework or in the A/B led one. The first A-B couple drawing is common to both of the settings: $n_F$ features (in the shown example $n_F=7$) are randomly picked for generating the control package and the variation $B$ is assembled varying the control. From the second drawing on, the two setting differ in the A-B couples way of selection. In the pure social spreading case, one of the two previous packages is chosen for the new round (new control $A$) and challenged against its varied version (new variation $B$). In the A/B led social spreading setting, the Bayesian A/B testing methodology is employed on the gathered click-through rates at the end of every round; in case the variation $B$ overperforms the control $A$ with a probability greater than a fixed threshold (in the shown simulation the threshold is set at $95\%$), then the successful $B$ becomes the new control $A$ for the next round and a new variation $B$ is generated varying the control. Al the contrary, if successful conditions for the A/B are not satisfied, the control $A$ remains the same and a new variation is generated from it. 
In every case, generating the variation depends on a mutation rate $\mu$(in the example shown, $\mu=3/7$).
The simulation of every scenario was performed for $R=100$ Monte Carlo replicas. Every replica was built according to a unique set of random seeds for pseudo-random generators; this guarantees either that replicas differ one another in random drawings employed or simulation reproducibility.

A crucial measure has been assessing the dynamics of the most successful linguistic features over time. We show a comparison of the two scenarios in Fig.\ref{fig:successful}. In a pure social spreading, the 'winning' linguistic feature behavior over time (averages over $100$ replicas) shows an evolution roughly fitted by a third degree polynomial function. At the contrary, in an A/B testing led scenario, the successful linguistic feature behavior follows a trend which can be fitted by a linear function. 
\vspace{5mm}
\begin{figure}[h]
\centering
\includegraphics[scale=0.3]{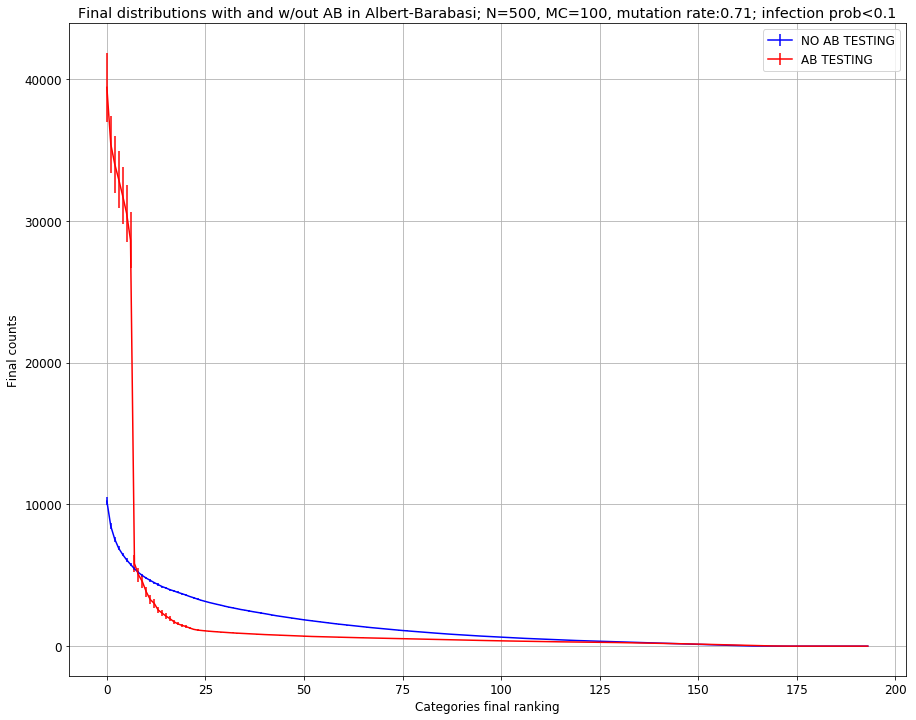}
\caption{Comparison of the final ranking distributions of messages in both explored scenarios, a pure social spreading setting and an A/B testing led one. The network structure employed in the simulation is the Albert-Barabasi one. }
\label{fig:distribution}
\end{figure}

\section{Conclusions}

This preliminary results coming up from this exploration are indicating that A/B testing has a substantial influence on the qualitative dynamics of information dissemination on a social network. Through measuring homogeneity and heterogeneity of final distribution of successful messages in a social network framework, either in a pure social spreading or in a AB testing-leaded one, we clearly observe (Fig.\ref{fig:distribution} ) that the A/B testing-led framework changes the dynamics of the pure social spreading qualitatively, by drawing crowd attention on leading linguistic features. In other words, we observe that A/B testing performed on synthetic social networks structure kills message heterogeneity by promoting the most successful linguistic features identified during the dynamics.
In the subsequent data analysis, leveraging the full, confirmatory upworthy data set, we will be able to quantify specifically which of the linguistic features are the ones particularly favoured by A/B testing strategies and allow inferences about its potential impact on the online discourse. Due to the dramatic shifts of the public discourse to algorithmically driven platforms, this assessment is of interest for the future design of platforms and regulation that preserve an exchange of arguments while content delivery stays relevant.

\bibliography{bibabtesting} 
\bibliographystyle{chicago} 
%\printbibliography

\end{document}